\documentclass[12pt,preprint]{aastex}
\newcommand{\COBE}{{\sl COBE}}
\newcommand{\ee}[1]{\ensuremath{\times 10^{#1}}}
\newcommand{\redchisq}{\ensuremath{\chi^2/\mathrm{d.o.f}}}
\newcommand{\secref}[1]{Section \ref{#1}}
\newcommand{\mum}{\ensuremath{~\mu \mathrm{m}}}
\newcommand{\kel}{\ensuremath{~\mathrm{K}}}

\newcommand{\chonebc}{175}
\newcommand{\chtwobc}{245}
\newcommand{\chthreebc}{400}
\newcommand{\chfourbc}{460}
\newcommand{\chfivebc}{630}
\newcommand{\dirbetenbc}{1250}
\newcommand{\dirbeninebc}{2140}
\newcommand{\dirbeeightbc}{3000}
\newcommand{\dirbesevenbc}{5000}

\def\Figure#1#2#3{
\begin{figure}[htb]
\epsscale{1.0}
\plotone{#1}
\caption{#2}
\label{#3}
\end{figure}
}

\shorttitle{Millimeter Spectrum of the Magellanic Clouds}
\shortauthors{Aguirre et al.}

\begin{document}

\title{The Spectrum of Integrated Millimeter Flux of the Magellanic Clouds 
and 30-Doradus from TopHat and DIRBE Data}

\author{J.~E.~Aguirre, \altaffilmark{1}
J.~J.~Bezaire, \altaffilmark{1}
E.~S.~Cheng, \altaffilmark{2}
D.~A.~Cottingham, \altaffilmark{3}
S.~S.~Cordone, \altaffilmark{4}
T.~M.~Crawford, \altaffilmark{5}
D.~J.~Fixsen, \altaffilmark{6}
L.~Knox, \altaffilmark{7}
S.~S.~Meyer, \altaffilmark{1,5}
H.~U.~Norgaard-Nielsen, \altaffilmark{8}
R.~F.~Silverberg, \altaffilmark{9}
P.~Timbie, \altaffilmark{4}
G.~W.~Wilson \altaffilmark {10}}

\altaffiltext{1}{University of Chicago,
                 Department of Physics, 
                 5640 South Ellis Avenue,
                 Chicago, IL  60637}
\altaffiltext{2}{Conceptual Analytics, LLC, 
                 8209 Woburn Abbey Road,
                 Glenn Dale, MD 20769}
\altaffiltext{3}{Global Science and Technology, Inc.,
                 Laboratory for Astronomy and Solar Physics, 
                 NASA / Goddard Space Flight Center, Code 685,
		 Greenbelt, MD 20771}
\altaffiltext{4}{University of Wisconsin at Madison,
                 Department of Physics, 
                 1150 University Avenue,
                 Madison, WI  53706}
\altaffiltext{5}{University of Chicago, 
                 Department of Astronomy,
                 5640 South Ellis Avenue,
                 Chicago, IL  60637}
\altaffiltext{6}{SSAI, 
                 Laboratory for Astronomy and Solar Physics,
                 NASA / Goddard Space Flight Center, Code 685, 
		 Greenbelt, MD 20771}
\altaffiltext{7}{University of California, 
                 Department of Physics, 
                 One Shields Avenue,
                 Davis, CA  95616}
\altaffiltext{8}{Danish Space Research Institute,
                 Juliane Maries Vej 32, 
                 DK-2100 Copenhagen, 
                 Denmark}
\altaffiltext{9}{NASA / Goddard Space Flight Center, 
                 Laboratory for Astronomy and Solar Physics, Code 685, 
                 Greenbelt, MD 20771}
\altaffiltext{10}{University of Massachusetts,
                 Department of Astronomy,
                 619E LGRT-B,
                 710 North Pleasant Street,
                 Amherst, MA 01003-9305}

\begin{abstract}
We present measurements of the integrated flux relative to the local
background of the Large and Small Magellanic Clouds and the region
30-Doradus (the Tarantula Nebula) in the LMC in four frequency bands
centered at \chtwobc, \chthreebc, \chfourbc, and \chfivebc ~GHz, based
on observations made with the TopHat telescope.  We combine these
observations with the corresponding measurements for the DIRBE bands
8, 9, and 10 to cover the frequency range \chtwobc~- \dirbeeightbc
~GHz ($100 - 1220 \mum$) for these objects.  We present spectra for
all three objects and fit these spectra to a single-component greybody
emission model and report best-fit dust temperatures, optical depths,
and emissivity power-law indices, and we compare these results with
other measurements in these regions and elsewhere.  Using published
dust grain opacities, we estimate the mass of the measured dust
component in the three regions.

\end{abstract}


\section{Introduction}
\label{sec:Introduction}

There has been a recent upsurge in interest in interstellar dust
emission, both in our own galaxy and in extragalactic environments.
Galactic dust emission in the far-infrared (FIR) and microwave is a
primary source of contamination to measurements of the Cosmic
Microwave Background (CMB)
\citep{masi01,jaffe03}.  
Association of the observed Cosmic Infrared Background (CIB) with dust
emission from high-redshift galaxies
\citep{puget99,scott2000} 
and the recognition of the potential use of this source as a probe of
structure formation
\citep{guiderdoni98,blain99a,haiman00,kceh01} 
has placed a high priority on understanding extragalactic dust
properties.

Most of what we know about interstellar dust from our own and other
galaxies comes from measurements of extinction (absorption plus
scattering) in the ultraviolet (UV), optical, and near- to
mid-infrared (NIR, MIR)
\citep{mathis90}.  
Extinction measurements can be compared to dust models
\citep{mathis77,drainelee84,li01}
to constrain the size, composition, and density of dust grains.
Low-frequency ($\nu \ll 300$ THz) dust emission is thought to be
dominated by thermal greybody emission from dust grains heated by the
interstellar radiation field (IRF), so to turn the knowledge of the
dust from extinction into a prediction for low-frequency dust
emission, one needs two further pieces of information: the spectrum of
the IRF and the greybody emissivity of the different grain populations
--- in general a function of frequency, grain size, and grain
composition.  Conversely, one can use measurements of low-frequency
dust emission combined with models of the dust density and optical
properties to constrain models of the IRF.

In the simplest models of optical properties of dust at low
frequencies, the dust emissivity is independent of grain size and
composition.
\citet{hildebrand83} 
and
\citet{drainelee84} 
(hereafter DL84) argue that for dust grains much smaller than a
wavelength ($\lambda = c/\nu \gg a$, where $a$ is the grain radius),
the emission cross-section will be proportional to the volume of the
grain.  The emissivity per unit dust mass for a given dust density is
then independent of the nature of the grain size distribution as long
as all grains are small compared to the wavelengths of interest.  DL84
argue further that electric dipole radiation will dominate in this
limit, and the dust emissivity will be proportional to the square of
the frequency, regardless of dust composition.  Later theoretical work
\citep{tielens87} 
and laboratory measurements
\citep{agladze96} 
have suggested different power-law indices for IR dust emissivity, 
depending on composition, and ranging from $\nu^1$ to 
$\nu^{2.7}$, even in the regime $\lambda = c/\nu \gg a$.
\citet{agladze96} have also found evidence for 
temperature-dependent behavior of the power-law index for certain
types of amorphous silicate grains.

Further complicating matters is the possibility that different dust
populations along a single line of sight can have different emission
temperatures, even if the radiation field does not change along the
line of sight.  Grains in the diffuse interstellar medium (ISM) are
predicted to have temperatures of $\sim 10 - 20 \kel$ (DL84), but very
small grains can be transiently heated to temperatures of hundreds of
Kelvin by a single UV photon and will cool to ambient temperature by
re-emitting in the NIR/MIR
\citep{sellgren84}.  
\citet{li01} 
estimate the maximum grain radius $a$ at which this effect is
noticeable to be $a \sim 25~\mathrm{nm}$.  They divide silicate and
carbonaceous (graphite) dust into ``big'' and ``small'' populations
based on this criterion and estimate emission from small grains to be
an important contributor at $5000$ GHz but small compared to
large-grain emission at $3000$ GHz and lower frequencies.

\citet{dunne01} 
have combined data from the Infrared Astronomy Satellite ({\sl IRAS})
at $5000$ GHz ($60 \mum$) and $3000$ GHz ($100 \mum$) with $670$ GHz
($450 \mum$) and $350$ GHz ($850 \mum$) data from the Sub-millimetre
Common User Bolometer Array (SCUBA) camera and other sub-mm
measurements in the literature to characterize the dust emission from
32 nearby galaxies.  In an earlier study,
\citet{dunne00} 
found that fitting just the two {\sl IRAS} points and the SCUBA $350$
GHz point resulted in derived temperatures of $35.6 \pm 4.9 \kel$ and
emissivity indices of $1.3 \pm 0.2$ for a larger sample of 104 nearby
objects.  The new study finds that for the sources for which they were
able to add more sub-mm points, the spectra were better fit by a model
with two components at different temperatures, with $31 \kel < T_{hot}
< 60 \kel$ and $18 \kel < T_{cold} < 32 \kel$.  The authors note that
these conclusions are not definitive, due both to a paucity of
measurements for most of their sources (only 10 of their sources are
measured in more than four spectral bands --- only one in more than
six bands --- and a two-component greybody model with power-law
emissivity in general has six free parameters) and to a fundamental
degeneracy between a broadened temperature distribution and a
shallower emissivity power-law.

In this paper we consider high-fidelity measurements in seven
frequency bands of the two galaxies nearest to ours: the Large and
Small Magellanic Clouds.  All of the measurements are at frequencies
$\nu \le 3000$ GHz.  The measurements at $\nu \ge 1250$ GHz are taken
from the publicly available data from the Diffuse Infrared Background
Experiment (DIRBE) on the Cosmic Background Explorer ({\sl COBE})
satellite, while the measurements at $\nu < 1250$ GHz are new results
from the TopHat instrument.

The irregular dwarf galaxies known as the Magellanic Clouds are the
most prominent extragalactic features in the southern sky and have
been observed for hundreds of years in many spectral bands (for a
summary, see
\citet{westerlund97} 
or
\citet{vandenbergh00}).  
Despite the proximity of these two galaxies, there exist relatively
few observations of them in the FIR or in continuum sub-millimeter or
microwave bands.

Studies of dust in the Magellanic Clouds have been done using
extinction measurements
\citep{rodrigues97,gordon98,misselt99,weingartner01}.
\citet{weingartner01} estimate dust grain size distributions in the
Magellanic Clouds and our galaxy and find that the distribution shape
and large-radius cutoff are similar in the two environments.  The
product of grain size distribution and emissivity per grain increases
with grain size in the two types of grains in the Weingartner \&
Draine model (silicates and carbonaceous (graphite) grains) up to a
cutoff at $a \sim 0.2 \mum$ for silicates and $a \sim 1 \mum$ for
graphites.  Based on this work, we assume grains large enough to be in
thermal equilibrium with the IRF will dominate the emission at $\nu
\le 3000$ GHz from the Magellanic Clouds.

Attempts have been made to characterize the dust in the Magellanic 
Clouds using {\sl IRAS} data alone, 
\citep{sauvage90,stanimirovic00}, while 
\citet{andreani90}
combined {\sl IRAS} data with single scans across the LMC and SMC with a
ground-based millimeter-wave receiver operating from Antarctica.  {\sl IRAS}
observed the entire sky at $5\arcmin$ resolution in four spectral
bands centered at $25000$, $12000$, $5000$, and $3000$ GHz 
(12, 25, 60, and 100\mum).  However, {\sl IRAS} was
designed to detect point sources, and though large-area sky maps have
been created from {\sl IRAS} data (ISSA images,
\citet{wheelock93}), 
large zero-point and calibration drifts across these maps render them 
unsuitable for determinations of absolute flux in regions as large as 
the Magellanic Clouds.

DIRBE extended the frequency range of {\sl IRAS} with two lower-frequency
channels ($2140$ GHz ($140 \mum$) and $1250$ GHz ($240 \mum$)) and was 
specifically designed to
measure diffuse emission, so the zero-point and absolute calibration
are more well-behaved and well-characterized.  The angular resolution
of DIRBE is $0.7\arcdeg$, which is well-matched to characterizing the
integrated dust properties of the Magellanic Clouds, which have
angular extents on the order of degrees.
\citet{schlegel98}
(hereafter SFD98) have combined {\sl IRAS} and DIRBE $3000$ GHz
(100\mum) data to produce a full-sky model with $5\arcmin$ resolution
and the stability of the DIRBE data.
\citet{finkbeiner99} 
(hereafter FDS99) use DIRBE $1250$ GHz data and data from the {\it
COBE} Far Infrared Absolute Spectrophotometer (FIRAS) to extend this
model to longer frequencies, but they exclude the Magellanic Clouds
from the extended model.
\citet{stanimirovic00} 
and
\citet{li02} 
use DIRBE data and the SFD98 model to characterize the dust in the
SMC, and
\citet{stanimirovic00} 
find that some combination of components at temperatures between $15
\kel$ and $30 \kel$ and an emissivity proportional to $\nu^2$ fits
the combined {\sl IRAS}/DIRBE/SFD98 data at $\nu \le 5000$ GHz.  Similar
analyses have not been published for the integrated dust emission from
the LMC.

We report here a measurement of the spectra of integrated flux
relative to the background of the LMC and SMC.  We report separate
results for the LMC with the active star-forming region 30-Doradus
(the Tarantula Nebula) masked off and for 30-Doradus alone.  This
measurement is based on observations made with TopHat and DIRBE, and
spans the range \chtwobc~-\dirbeeightbc ~GHz.  We report calibrated
spectra (in Jy, $1 \; \mathrm{Jy} = 10^{-26} \; \mathrm{W}
\mathrm{m}^{-2} \; \mathrm{Hz}^{-1}$) of all three regions.  We fit the
calibrated spectra to a single-component greybody emission model with
power-law emissivity and report a best-fit optical depth, temperature,
and emissivity power-law index for each region.  These new
measurements provide data which span a gap in our knowledge of the LMC
and SMC integrated spectra.  The dust emission properties are
complicated by neither stochastic, non-equilibrium dust grain heating
nor resonance emission, which are important factors at frequencies
above 3 THz. In addition, the sub-mm measurements in the
Rayleigh-Jeans portion of the dust emission permit an estimation of
the dust mass.  In particular, if a steep emission spectrum is
indicated for theoretical reasons, a large mass of cold gas would be
required to match the flat sub-mm spectrum reported here.

The paper is arranged as follows.  
\secref{sec:Instrument} 
describes the TopHat instrument and observations and
\secref{sec:DataAnalysis} 
describes the reduction of raw TopHat data to uncalibrated sky maps
and uncertainties.  
\secref{sec:FluxAnalysis} 
describes the flux analysis performed and the regions selected.
\secref{sec:DIRBE} 
explains how we treat the DIRBE data and combine it with the TopHat
observations.  The TopHat calibration and the effect of the DIRBE
calibration uncertainties are discussed in
\secref{sec:Calibration}, 
and the final calibrated fluxes with errors are given in
\secref{sec:CalibratedFluxes}.  
We discuss the physical interpretation of these results and compare to
results from the literature in
\secref{sec:Discussion},
and we discuss applications of these results and future work in
\secref{sec:Conclusions}.

\section{Instrument Description and Observations}
\label{sec:Instrument}

TopHat \citep{cheng93, silverberg03} is a balloon-borne telescope
designed to measure millimeter astrophysical emission over a large
area of sky.  TopHat is an on-axis Cassegrain telescope with a 1 m
aluminum primary and a secondary mirror suspended on six Kevlar
fibers.  The beam was designed to be an approximately $20\arcmin$ FWHM
top hat, and ground measurements of the beam profile were consistent
with the predicted shape.  In addition, ground measurements were
performed of the far sidelobe response, and the rejection was measured
to be $> 80$~dB at angles greater than $25\arcdeg$ from the optical
axis, and $> 110$~dB at angles greater than $70\arcdeg$.  The
instrument is mounted on top of a scientific balloon, with a gondola
providing support electronics, power, and telemetry hanging beneath.
The telescope and a conical radiation shield are mounted on a rotating
azimuthal mount, with the optical axis fixed at a $12
\arcdeg$ inclination from the rotation axis.  Observations are made by
rotating the mount at a constant rate of one rotation per 16 sec.
Observing at
\mbox{$78\arcdeg$ S}
latitude, the scan pattern as the earth rotates becomes a series of
interlocked circles whose centers circumscribe the South Celestial
Pole (SCP) once each sidereal day.  The nominal observed region is
then approximately a
\mbox{$48\arcdeg$}
diameter circle, centered on the SCP.  The actual scan pattern
obtained was more complicated and covered a slightly larger area due
to a tilt of the telescope from horizontal and the rotation of the
balloon.

The instrument observes a single pixel in five spectral bands, each
with a single detector.  The band is defined by an IR absorber,
resonant grid beam splitters, and band defining filters.  In this
work, we define the center frequency of the band $\langle \nu
\rangle$ as an effective Rayleigh-Jeans (RJ) frequency by
\begin{equation}
\langle \nu \rangle = 
\left[ \frac{\int{\nu^2 \; t(\nu) \; d\nu}}
{\int{t(\nu) \; d\nu}} \right]^{1/2}
\end{equation}
where $t(\nu)$ is the transmission of the band.  The center
frequencies are then \chonebc, \chtwobc, \chthreebc, \chfourbc, and
\chfivebc ~GHz, with $\Delta \nu / \nu \sim 25\%$ in the two 
lowest-frequency channels and $\sim 10\%$ in the highest three.  The
spectral bandpasses were determined before the flight using Fourier
transform spectrometry; the bandpasses of the four highest-frequency
bands, which are used in this analysis, are shown in Figure
\ref{fig:bandpass}.
The width of the bands leads to appreciable color corrections for
source spectra that differ appreciably from RJ.  These corrections are
incorporated into spectral fits as described in Section
\ref{sec:Calibration}; the calculation of the corrections themselves 
(for both TopHat and DIRBE) is discussed in Appendix 
\ref{app:ColorCorrections}.  
The detectors are five silicon bolometers with ion-implanted
thermistors cryogenically cooled to 270~mK by a $^3$He cryostat; the
dewar and its internal electronics are described in detail in
\citet{fixsen01} 
and 
\citet{oh01}.  
The band at \chonebc ~GHz is not used in this analysis because of
excess noise.

\clearpage

\Figure
{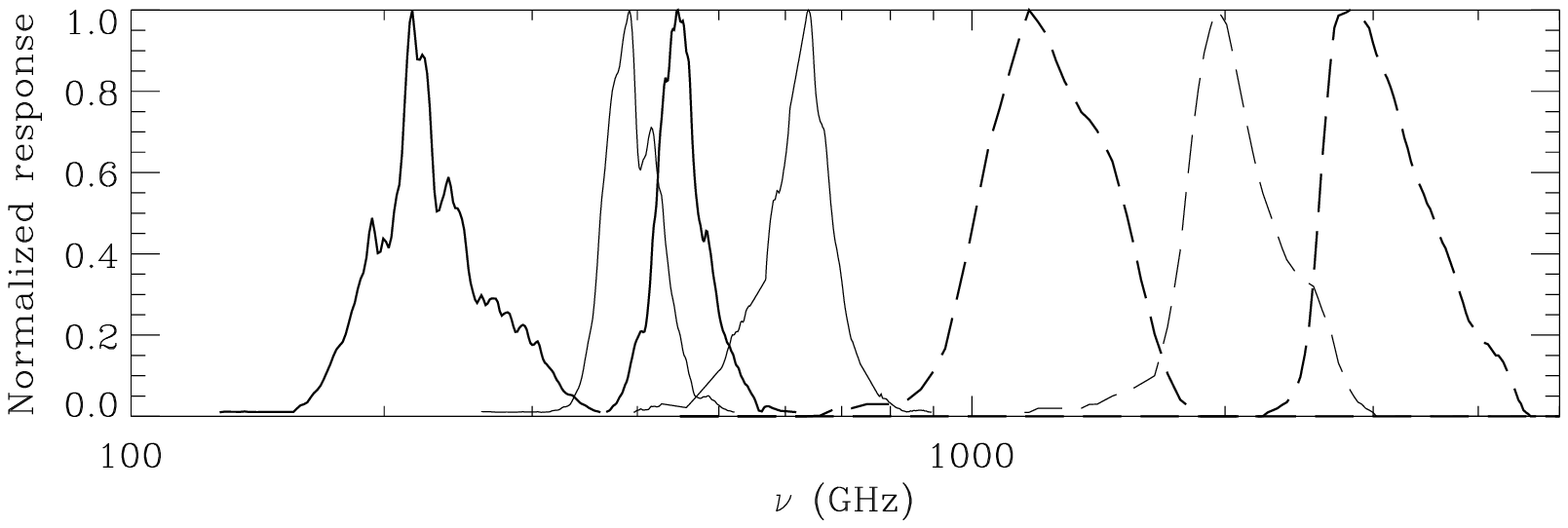}
{Bandpass relative to a white spectrum of the four TopHat bands used
in this analysis, together with DIRBE bands 8, 9, and 10.  The TopHat
bands are shown in alternating heavy and light solid lines, and the
DIRBE bands in heavy and light dashed lines.  The transmission of each
band has been normalized to unity at the peak.}  {fig:bandpass}

\clearpage

The telescope is not actively pointed.  The location of the beam is
reconstructed after the flight using a two-axis tilt meter, GPS
information, and the location of the sun as determined by four small
imaging telescopes on the radiation shield of the telescope, which
record the sun position four times each rotation.  A pointing model
was constructed which uses the data from these sensors and depends on
a small number of parameters to locate the spin axis and the beam.
The parameters are determined by a multi-dimensional fit of the
observations containing the Galaxy to a map.  The observations of the
Magellanic Clouds are not used in the pointing fit; rather, we take
the reproducibility of their location on repeated observations as
independent evidence of the quality of the pointing model.  The RMS
pointing errors are determined to be $\sim 3 - 4 \arcmin$, much
smaller than the beam, and are negligible for this analysis.

TopHat was launched from McMurdo Station, Antarctica at 06:55 UT 2001
January 4 by the National Scientific Balloon Facility (NSBF).  After
an initial checkout, sky observations began at 14:10 UT 2001 January 4
and continued until 14:00 UT 2001 January 8 when the cryogens were
exhausted.  All observations occurred at a float altitude of $37.5 \pm
1.4$~km.  The telescope was turned off on 2001 January 10, and the
balloon cut down and the flight disks recovered on 2001 January 31.
The observations consist of a nearly circular patch of sky centered at
the South Celestial Pole (SCP), approximately $30\arcdeg$ in radius,
or about 6\% of the sky, re-observed four times over the course of
$\sim$4 sidereal days.  Because of interruptions in the observations
for system checks, the data exists in only three continuous sections.
In addition, data taken during the first 16 hours of observation
proved unusable due to long-settling transients in many of the
systems.  We have divided the remaining observations into two
independent sets, denoted Epoch I and Epoch II, with approximately
equal observing weight, so that all steps in the analysis may be
performed on both epochs and the results compared.  These divisions
are summarized in Table
\ref{tab:ObservingTimes}.  

\clearpage

\begin{deluxetable}{cll}
\tablewidth{0pt}
\tablecaption{Observing times of data used in this analysis.}
\tablehead{
\colhead{Epoch}       & 
\colhead{Start}       & 
\colhead{End}
}
\startdata
\label{tab:ObservingTimes}
I  & 04:18 UT 2001 Jan 5 & 15:28 UT 2001 Jan 5 \\
   & 15:43 UT 2001 Jan 5 & 00:33 UT 2001 Jan 6 \\
   & 17:55 UT 2001 Jan 7 & 14:00 UT 2001 Jan 8 \\
\tableline
II & 00:33 UT 2001 Jan 6 & 17:28 UT 2001 Jan 6 \\
   & 18:03 UT 2001 Jan 6 & 17:55 UT 2001 Jan 7 
\enddata
\end{deluxetable}

\clearpage

\section{Data Analysis}
\label{sec:DataAnalysis}

The raw radiometer data are reduced by constructing a minimum variance
fit of the timestream to a sky map.  Prior to mapmaking, the data are
processed as follows.  Cosmic ray strikes and other anomalies are
removed from the time stream data and the instrument transfer function
is deconvolved.  Then instrumental effects in the data are identified;
large instrumental signals of various kinds which are localized in
time are excised and ignored in further analysis, while
scan-synchronous instrumental signal is dealt with by fitting the
timestream simultaneously to the sky and a model of this
scan-synchronous signal.  and contaminated data either cut or an
appropriate model constructed by which the contaminant may be removed
in a simultaneous fit.

The
timestream noise is estimated iteratively from a series of
intermediate fit residuals.  A model of the telescope pointing is
constructed and its parameters fit to minimize the map errors.  The
final sky map is a vector $m$, produced by a linear fit to the
accepted data in the cleaned timestream $d$ which minimizes the
$\chi^2$ function
\begin{equation}
\label{eq:mapchisq}
\chi^2 = (d - A(p) m)^T W (d - A(p) m)
\end{equation}
where $d$ is the cleaned data timestream, $A$ is the pointing matrix,
a function of the pointing parameters $p$, which is supplemented by
model templates to account for instrumental effects, and $W$ is the
inverse of the time-time noise covariance matrix, $W = \langle n n^T
\rangle^{-1}$.  The pointing matrix is pixelized using 
HEALPix\footnote{{\tt http://www.eso.org/science/healpix/}}
\citep{gorski99} 
with pixels $14 \arcmin$ on a side.  The best fit map is given by
\begin{equation}
\label{eq:map}
m = (A^T W A)^{-1} A^T W d = N A^T W d
\end{equation}
where the covariance matrix $N$ for this simultaneous estimation of
sky pixels plus instrumental model parameters is given by
\begin{equation}
\label{eq:noisecovariance}
N = (A^T W A)^{-1}
\end{equation}
Equations 
\ref{eq:map} 
and 
\ref{eq:noisecovariance} 
are solved directly using MADCAP
\citep{borrill99}.  
Maps are produced for both Epochs separately.  In the following, we
marginalize $N$ over sky pixels which do not lie in the fields of
interest and over all instrumental parameters.  

\section{Flux Analysis}
\label{sec:FluxAnalysis}

TopHat has no absolute reference for the power coming from the sky; it
is sensitive only to flux differences between different positions.
Our formal uncertainty on any quantity with a nonzero projection onto
a constant sky flux is infinite; therefore we must construct a purely
differential quantity to estimate.  The integrated flux from an
extended source minus the integrated flux from a surrounding set of
pixels with the same total solid angle as the source is such a
quantity.  This approach is particularly appropriate in the case that
the flux from the on-source pixels is a combination of the source
emission and emission from a relatively uniform, optically thin
foreground; then the differential measurement gives the flux from the
source only.

There is some evidence of systematic contamination above the random
noise level in the TopHat data, particularly in spatial modes which
are a function of declination (Dec) only.  A large spin synchronous
instrument signal is fit out of each channel's timestream
simultaneously with the best-fit map.  While the pixel-pixel
correlations induced by this instrument model are included in the pixel
covariance matrices used in the flux analysis, the model lacks
sufficient fidelity to remove the spin-synchronous signals completely,
leaving a small residual spin synchronous signal improperly subtracted
and unaccounted for in the covariance matrix.  The symmetry of our
scanning strategy naturally projects any residual spin synchronous
signal mainly onto spatial modes in the map that are functions of
declination only.  For this reason, we have constructed the on- and
off-source regions to have the same number of pixels on each
iso-latitude ring in the HEALPix ``Ring'' pixelization, rendering the
difference between on- and off-source flux insensitive to these
potentially contaminated modes.  We arrange the off-source pixels
symmetrically around the on-source pixels so that the differential
flux is also insensitive (to first order) to other long-wavelength
modes that we do not wish to include in this measurement, such as the
CMB dipole.  The diameters of the on-source regions (chosen to be
circular for convenience) were optimized for the ``target'' Magellanic
Cloud regions by varying the diameter around an initial value selected
by eye to enclose the particular high-contrast area of the DIRBE
\dirbetenbc~GHz map and noting when the slope of the enclosed flux
versus diameter tends to zero.  This process is not used to optimize
the 30-Doradus on-source region because it is itself embedded in a
high-contrast region.  The diameter of the 30-Doradus on-source region
is chosen by eye.

Having chosen the source region and a background region satisfying the
above conditions, we then extract from the full map $m$ and covariance
matrix $N$ (Equations
\ref{eq:map}
and
\ref{eq:noisecovariance}) 
the pixels corresponding to the on and off-source regions.  This is
done separately for the two Epochs.  We denote the vector of extracted
pixel values $v$, and the matrix of extracted pixel-pixel covariances
$C$.  The weighting vector $w$ is defined such that it is 1 for the
on-source pixels and -1 for the off-source pixels.  We then calculate,
for each channel, the quantity
\begin{equation}
S = w^T v
\end{equation}
with a corresponding error variance of
\begin{equation}
\sigma^2_S = w^T C w
\end{equation}
We demonstrate the insensitivity of the weighting vector $w$ for the
LMC, SMC, and 30-Doradus regions to low spatial frequency modes by
giving the normalized projections\\ $(w^T\xi)/\sqrt{(w^T w) (\xi^T
\xi)}$ of various modes $\xi$ onto the weight vector $w$ in Table
\ref{tab:LowFreqProj}.  

\clearpage

\begin{deluxetable}{lrrrr}
\tablewidth{0pt}
\tablecaption{
Normalized projections of low spatial frequency modes onto the pixel
weight vectors used in the differential flux measurements.}
\tablecomments{
The ``cross dipole'' modes are the two dipole directions orthogonal to
the CMB dipole.  Cross dipole 2 points primarily along the direction
towards the SCP.}
\tablehead{
\colhead{Mode} & \colhead{LMC} & 
\colhead{30-Dor} & \colhead{SMC} & \colhead{Blank}
}
\startdata
\label{tab:LowFreqProj}
Constant offset & 0 & 0 & 0 & 0 \\ 
CMB dipole & -0.0093 & 0.0017 & 0.0147 & 0.0071 \\
Cross dipole 1 & 0.0136 & 0.0012 & 0.0079 & 0.0126 \\ 
Cross dipole 2 & 0.0002 & $< 1 \times 10^{-4}$ & 0.0027 & 0.0003 \\ 
Constant dec mode & 0 & 0 & 0 & 0 \\
\enddata
\end{deluxetable}

\clearpage

There are no sufficiently bright point sources between \chtwobc~and 
\chfivebc~GHz in the TopHat observing region, so we constrain the 
possible in-flight beam profile by a combination of ground
measurements and limits on how the optical configuration could
possibly change during launch and at altitude.  However, in this
analysis we have chosen the conservative approach of making all source
regions large compared to an upper limit on the beam size determined
by assuming the beam is the size of the most compact bright sources we
can find in our maps.  These most compact sources are no more than one
degree FWHM, and all source regions in this analysis are at least two
degrees in diameter.  Because of the very high sidelobe rejection of
the telescope, the contribution of the brightest off-axis sources (the
sun and the Galaxy) are negligible.

In addition to the source regions of the LMC, SMC, and 30-Doradus, we
have selected five regions of appreciable Galactic dust emission which
are out of the Galactic plane and one ``blank'' region on which to
perform the flux analysis.  The Galactic dust regions are used in the
calibration, discussed in
\secref{sec:Calibration}.  
These regions are numbered 1-5, with regions 3, 4, and 5 corresponding
to the Chameleon Nebulae.  The blank region is chosen such that its
differential brightness is consistent with zero in the DIRBE $1250$
GHz channel.  The TopHat flux of the blank region serves as a
consistency check on the method, since we expect to be dominated by
thermal dust emission correlated with DIRBE in the TopHat bands.

The on-source fields are all circular about the center given; their location
and the solid angle they subtend are given in Table
\ref{tab:regions}.  
The on- and off-source fields for each region are shown in
Figure
\ref{fig:regions}.

\clearpage

\begin{deluxetable}{lccccccc}
\tablewidth{0pt}
\tablecaption{Locations of regions used in the flux analysis.}
\tablehead{
Region & \multicolumn{3}{c}{RA Center}   & \multicolumn{3}{c}{Dec Center}
        & \multicolumn{1}{c}{Solid angle} \\
        & \colhead{(h)} & \colhead{(min)} & \colhead{(s)} & 
          \colhead{(deg)} & \colhead{(min)} & \colhead{(s)}
        & \colhead{(ster)}
}
\tablecomments{
All coordinates are J2000.}
\startdata
\label{tab:regions}
Dust 1 & 7 & 7 & 52.1 & -78 & 50 &  6.4 & .01596 \\
Dust 2 & 19 & 30 & 0.0 & -80 & 7 & 11.0 & .01596 \\
Dust 3 & 11 & 4 & 24.7 & -77 & 32 & 56.8 & .00118 \\
Dust 4 & 12 & 49 & 5.5 & -79 & 56 & 10.6 & .00113 \\
Dust 5 & 12 & 54 & 0.0 & -77 & 10 & 53.0 & .00118 \\
Blank & 23 & 16 & 24.4 & -66 & 26 & 36.7 & .00644 \\
\tableline
LMC & 5 & 18 & 27.7 & -68 & 29 & 35.6 & .01481 \\
SMC & 0 & 52 & 15.5 & -72 & 56 & 31.6 & .00544 \\
30-Dor & 5 & 39 & 28.4 & -69 & 3 & 3.5 & .00120 \\
\enddata
\end{deluxetable}

\clearpage

\Figure
{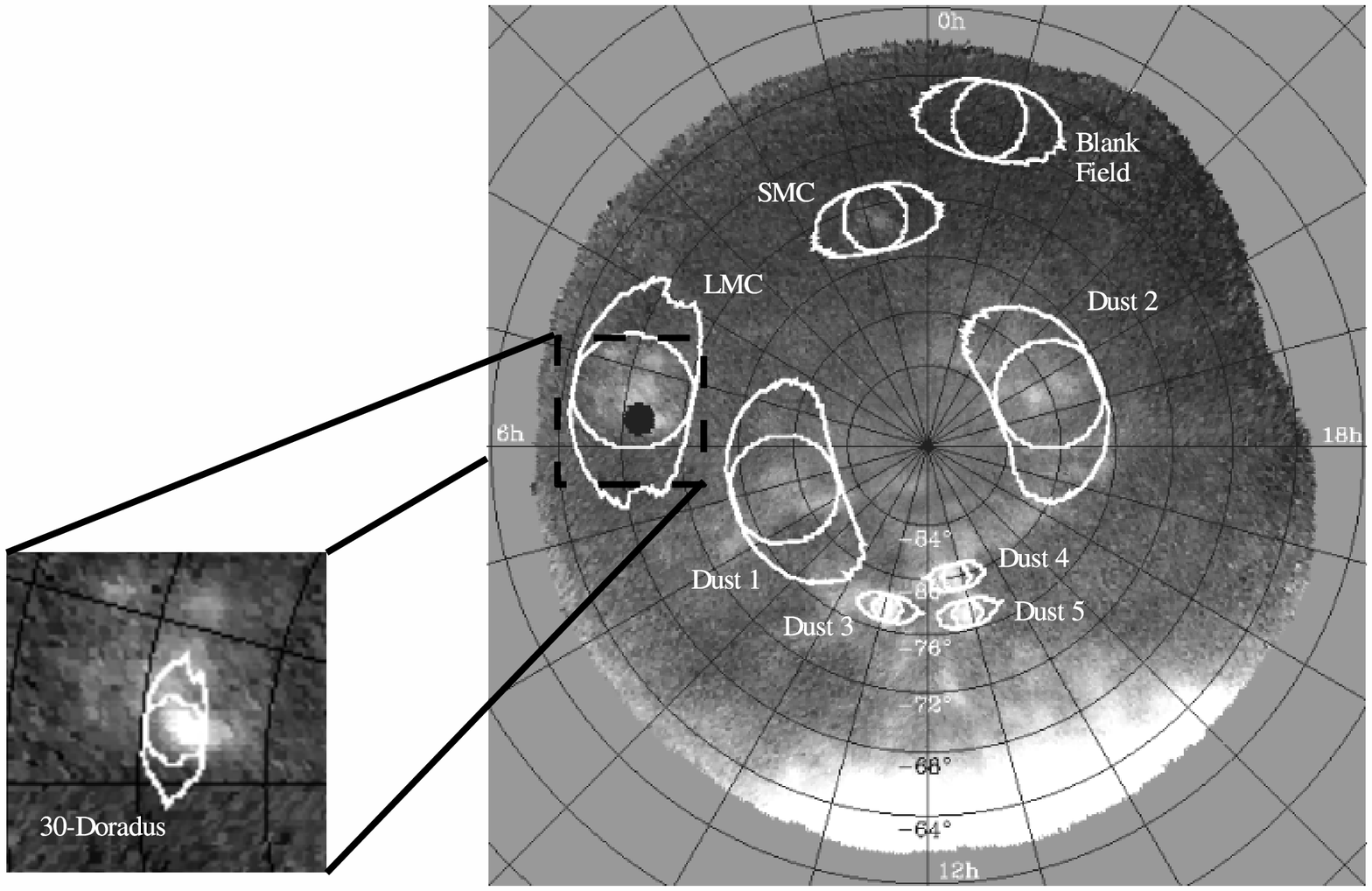}
{TopHat channel 5 combined Epoch I and II map showing the on-source
and off-source regions used in calculating the LMC, SMC, and blank
field flux, as well as the dust regions used for calibration.  The
region used for 30-Doradus is shown in the inset.  RA increases
counter-clockwise from the top in steps of 1 h per division; the SCP
is at the center of the map, with Dec increasing by 4\arcdeg~per
division.  In computing the flux of the LMC alone, the 30-Doradus
region (the central circle in the inset) is not included. }
{fig:regions}

\clearpage

We probe the consistency of the flux results obtained by several
$\chi^2$ tests.  We first ask, for a given Epoch, if the spectrum of
the blank field in all four TopHat channels is consistent with zero
flux.  The results of these tests are presented in Table
\ref{tab:NullChisq},
which gives the \redchisq~and the probability to exceed (PTE) this
value with a correct model.  We find that the combined TopHat
measurements of the blank region are consistent with zero in both
Epochs, but we note that this is not the case for an arbitrary choice
of off-source region shape, such as a region circularly symmetric
about the on-source pixels.  This choice of off-source region yields a
$\redchisq = 18.7/4$ for the model that the blank region is indeed
blank in Epoch II, with an associated PTE of 0.001.  We next ask if
the flux measurements are consistent between Epochs.  These results
for all regions selected are given in Table 
\ref{tab:ConsistentChisq}.
We find that the spectra measured are consistent between Epochs and
therefore in what follows we quote the fluxes from a weighted average
of the two Epochs.  Again, this conclusion is not reached with
off-source regions circularly symmetric about the on-source region.
For example, the LMC minus 30-Doradus has an Epoch-to-Epoch
consistency $\redchisq = 19.7/4$ with this naive choice of off-source
regions, with an associated PTE of $5 \times 10^{-4}$.  Finally, we
note that, as measured in the three DIRBE bands, the choice of
off-source region does not affect the measured differential flux above
the $\sim 1 \%$ level, except for the special case of 30-Doradus, for
which the different off-source regions sample different parts of the
still reasonably bright diffuse LMC.

\clearpage

\begin{deluxetable}{lcc}
\tablewidth{0pt}
\tablecaption{$\chi^2$ test of the null hypothesis for the blank field.}
\tablehead{
\colhead{Epoch} & \colhead{$\redchisq$} & \colhead{PTE}
}
\startdata
\label{tab:NullChisq}
I  &  5.165/4 & 0.27 \\
II &  1.438/4 & 0.84 \\
Sum & 6.603/8 & 0.58 \\
\enddata
\end{deluxetable}

\begin{deluxetable}{lcc}
\tablewidth{0pt}
\tablecaption{
Results of $\chi^2$ tests of consistency between TopHat Epochs for
observed fields.}
\tablehead{
\colhead{Region} & \colhead{$\redchisq$} & \colhead{PTE}
}
\startdata
\label{tab:ConsistentChisq}
Dust 1  &  8.556/4 & 0.07 \\
Dust 2  &  4.155/4 & 0.39 \\
Dust 3  &  4.803/4 & 0.31 \\
Dust 4  &  1.051/4 & 0.90 \\
Dust 5  &  4.344/4 & 0.36 \\
Blank   &  4.699/4 & 0.32 \\
\tableline
LMC     &  3.757/4 & 0.44 \\
SMC     &  3.202/4 & 0.52 \\
30-Dor  &  0.901/4 & 0.92 \\
\enddata
\end{deluxetable}

\clearpage

\section{Treatment of the DIRBE Data}
\label{sec:DIRBE}

Having demonstrated the internal consistency of the TopHat data, we
proceed to combine it with the DIRBE data set to extend the range of
the spectrum.  The DIRBE observations are the closest in frequency to
our own, extending up in frequency from \dirbetenbc ~GHz, and with a
beam of 0.7\arcdeg~they are well-suited to measurements of objects on
the angular scales of the Magellanic Clouds
\citep{kelsall98, hauser98}.
We use the Zodi-Subtracted Mission Average (ZSMA) intensity and
standard deviation data from bands 8, 9, and 10, with nominal band
centers of \dirbetenbc, \dirbeninebc, and \dirbeeightbc ~GHz.  The
bandpasses are shown in Figure
\ref{fig:bandpass}.
DIRBE reports their nominal band centers relative to source spectrum
with constant $\nu I_\nu$; the color corrections for a finite set of
other source spectra are available as part of the DIRBE Explanatory
Supplement 
\citep{des98}
or one can calculate corrections for an arbitrary spectrum from the
published DIRBE passbands, available electronically.\footnote{ {\tt
http://space.gsfc.nasa.gov/astro/cobe/dirbe\_exsup.html}} We
incorporate corrections for various model spectra in a spectral fit as
described in Section 
\ref{sec:Calibration}; 
the calculation of the corrections is discussed in Appendix
\ref{app:ColorCorrections}.  

The DIRBE maps exist in the {\sl COBE} quadcube pixelization, so for
the purposes of comparison it must be re-pixelized.  We re-pixelize
by resampling the quadcube pixelization onto the 14\arcmin~HEALPix
pixelization; the standard deviations per pixel are similarly
repixelized.  We note that the 14\arcmin~pixels used oversample the
DIRBE beam.  We then select the same regions as for the TopHat
analysis and perform the differencing of the integrated flux in an
identical fashion.  This allows us to combine the DIRBE measurements
straightforwardly with our own.  We note that for the differential
analysis we make, the errors in flux due to the DIRBE absolute offset
uncertainty and to zodiacal subtraction uncertainty are negligible.

The gain calibration errors, however, are not negligible and must be
considered in addition to the random errors given by the standard
deviation on the ZSMA maps.  Our estimate of the DIRBE gain
calibration error is based on the work of
\citet{hauser98} 
and the cross-calibration of the DIRBE gains using FIRAS
\citep{fixsen97}.
The cross-calibration with FIRAS at \dirbeninebc~and \dirbetenbc~GHz
resulted in slightly different best-fit DIRBE gains and improved error
bars on those gains.  The
\citet{hauser98} 
error at \dirbeeightbc ~GHz was not improved because half of the DIRBE
passband was outside the FIRAS spectral coverage.  The gains and
errors found by the various authors are given in Table 
\ref{tab:DIRBEGain}.
We describe how these gains and uncertainties are incorporated into
the analysis in
\secref{sec:Calibration}.

\clearpage

\begin{deluxetable}{lcccccc}
\tablewidth{0pt}
\tablecaption{DIRBE gains and uncertainties.}
\tablehead{\colhead{Reference} & \multicolumn{2}{c}{\dirbeeightbc~GHz} &
\multicolumn{2}{c}{\dirbeninebc~GHz} & 
\multicolumn{2}{c}{\dirbetenbc~GHz} \\
 & \colhead{Gain} & \colhead{Uncertainty} &
\colhead{Gain} & \colhead{Uncertainty} &
\colhead{Gain} & \colhead{Uncertainty}
}
\tablecomments{
All gains and uncertainties given are dimensionless and are referred
to the Hauser et al. 1998 values.}%\cite{hauser98} values.}
\startdata
\label{tab:DIRBEGain}
Hauser et al. 1998   & 1.00 & 0.135 & 1.00 & 0.106 & 1.00 & 0.116 \\
Fixsen et al. 1997   & 1.25 & 0.150 & 1.04 & 0.02  & 1.06 & 0.02  \\ 
This work, as prior  & 1.00 & 0.135 & 1.04 & 0.02  & 1.06 & 0.02  \\
This work, best fit  & 1.21 & $^{+.01}_{-.09}$     & 
                       1.03 & $^{+.02}_{-.00}$     &
                       1.05 & $^{+.02}_{-.00}$      \\
\enddata
\end{deluxetable}

\clearpage

\section{Calibration}
\label{sec:Calibration}

In the \chtwobc ~GHz channel, the CMB dipole is observed with high
signal-to-noise and is the dominant feature in that channel's map,
apart from the Galaxy, so it is used as a calibration source.  The
\chtwobc ~GHz calibration uses the best measurement of {\sl COBE}
\citep{fixsen02}
for the amplitude and direction of the dipole and applies a correction
due to the earth's rotation around the sun.  The error on the
calibration of the \chtwobc ~GHz channel is appreciably smaller than
the random errors in the flux measurements of this channel, so we
ignore this error in the following analysis.

In the higher-frequency channels, the CMB dipole is not detected with
significance, so another calibration must be used.  This is
complicated by the absence of known calibration sources in the range
\chthreebc~- \chfivebc ~GHz at angular scales of 1\arcdeg.
We therefore use an interpolation between the Galactic dust
measurements in the calibrated \chtwobc ~GHz channel and the
calibrated DIRBE channels at \dirbetenbc, \dirbeninebc, and
\dirbeeightbc ~GHz.  This calibration uses the fields labelled Dust 1
- 5 in Table
\ref{tab:regions}
and Figure
\ref{fig:regions},
and does not use the measurements of the ``target'' Magellanic Cloud
regions.

The spectral interpolation requires a model of this Galactic dust.  We
assume each of the dust regions is optically thin at all frequencies
of interest and has uniform dust temperature and optical properties
throughout, and we model the integrated flux from each region as
\begin{eqnarray}
\label{eq:DustModel}
F_\nu & = & \int d \Omega d \ell \; \rho_d(\Omega, \ell) 
\kappa_m B_{\nu}(T) \\
\nonumber & = & \Delta \Omega \; \tau_{\nu} \; B_{\nu}(T),
\end{eqnarray}
where $\tau_{\nu}$ is the mean optical depth along all lines of
sight through the source at frequency $\nu$, $B_\nu(T)$ is
the Planck blackbody brightness at frequency $\nu$ and
temperature $T$, $\kappa_m$ is the dust opacity in cm$^2$/g, $\rho_d$
is the dust mass density, and the integral is taken over lines of
sight and solid angle.  We assume a power-law emissivity to the dust
so that
\begin{eqnarray}
\label{eq:DustEm}
\tau_{\nu} & = &\int d \ell \; \rho_d(\ell) \; \kappa_m(\nu) \\
\nonumber & = &\int d \ell \; \rho_d(\ell) \; \kappa_m(\nu_0) 
(\nu/\nu_0)^{\alpha} \\
\nonumber & = & \tau(\nu_0) (\nu / \nu_0)^{\alpha},
\end{eqnarray}
where $\nu_0 = 600$ GHz.  

We then fit the measured flux from the five dust regions to the dust
model in Equation
\ref{eq:DustModel},
where each region is fit to its own spectrum with free parameters $T$,
$\alpha$, and $\tau(\nu_0)$, and each of the three unknown TopHat
calibrations and the three DIRBE calibrations is allowed to vary but
forced to be identical for the five regions.  We assume a Gaussian
prior on the DIRBE calibrations of the best combination of the
\cite{hauser98} 
and
\cite{fixsen97}
values, as given in Table
\ref{tab:DIRBEGain}.  
This leads to minimizing the $\chi^2$ function
\begin{equation}
\chi^2 = \sum_{i=1}^5 \sum_{j=1}^7 
\left( \frac{F(i,j) / K(i,j) - c(j) \bar{F}(i,j)}{\sigma_F(i,j)} \right) ^2 + 
\sum_{j=5}^7 \left( \frac{c(j) - \bar{c}(j)}{\sigma(c(j))} \right)^2,
\end{equation}
where $i$ runs over the five dust regions and $j$ runs over the seven
bands, $F(i,j)$ is the uncalibrated TopHat or DIRBE flux for region
$i$ in band $j$, $\bar{F}(i,j)$ is the model flux from that region in
that band, $c(j)$ is the calibration for that band (a free parameter
in every band but the $245$ GHz TopHat channel), $\sigma_F(i,j)$ is
the uncertainty on the uncalibrated flux in that region and band,
$\bar{c}(j)$ is the nominal value of the calibration in the three
DIRBE bands, and $\sigma(c(j))$ is the uncertainty on that value.  The
color correction $K(i,j)$ for band $j$, given the model spectrum for
region $i$, is the scaling that accounts for the fact that the
effective center frequency for a band with finite width will be
different depending on the assumed source spectrum.  The method for
computing the color corrections is explained in Appendix
\ref{app:ColorCorrections}.  
The fit is done iteratively, which allows us to apply a color
correction at each iteration derived from the source spectrum
determined in the previous iteration.  This gives a fit with 35 data
points and 21 parameters and thus 14 degrees of freedom.

The best fit to the data has \redchisq = 36/14, with most of the
excess $\chi^2$ coming from the TopHat points.  The fit residuals are
not obviously systematically distributed, implying that the high
\redchisq~may simply be due to an underestimation of the uncertainty,
rather than an inappropriate model.  We have tested this by performing
the calibration fit with a two-component dust model, and find that the
\redchisq~is not improved significantly by the additional parameters, 
which is consistent with the contention that the poor fit is not due
simply to an inadequate model. As we noted in
\secref{sec:FluxAnalysis}, 
there is some evidence of contamination in the maps which is not
properly accounted for in the noise covariance matrix, and while we
have mitigated this by our choice of regions in the flux analysis and
tested for noise misestimation across Epochs (Table
\ref{tab:ConsistentChisq}),
there may yet be excess residual contamination which becomes evident
when we attempt to combine arbitrary, widely separated regions in the
map, as the calibration fit attempts to do.  We therefore believe it
is justifiable, and conservative, to increase the TopHat uncertainty
estimation to account for this discrepancy.  We increase the TopHat
uncertainty estimation so that the TopHat contribution to the $\chi^2$
is comparable to that of the DIRBE points, which is accomplished by
doubling the TopHat errors.  This improves the
\redchisq~for the one-component dust model calibration fit to 20/14,
which has a PTE of 0.13.  We conclude from this that a one-component
model is an adequate model for the spectra observed, given the best
fit gains, if the estimate of the uncertainty of the TopHat fluxes is
increased.

The dust model parameters obtained for each region are given in Table
\ref{tab:DustParams}.
We note that the assumption of optical thinness of each region is
justified {\it a posteriori} by these results.  Table
\ref{tab:DustParams} 
also gives the square root of the appropriate diagonal element of the
covariance matrix for the parameters, which uses the increased TopHat
error estimate.  This is equivalent to the formal $1 \sigma$ error
with marginalization over all other parameters.  We stress that the
parameters have significant correlations between them.  The best fit
model for each region, along with its residual, is plotted in Figure
\ref{fig:ddcal}.
%The DIRBE calibrations preferred by this fit are given in Table
%\ref{tab:DIRBEGain}.
%We note that the band 9 and 10 best fit gain values are expected to be
%consistent with both
%\citet{hauser98} 
%and
%\citet{fixsen97}
%because of the prior assumption, and in fact are.  The band 8 value,
%however, is consistent with
%\citet{fixsen97} 
%but marginally inconsistent with
%\citet{hauser98}, 
%in spite of the assumption of a prior to the contrary; this
%determination is independent of the
%\citet{fixsen97}
%analysis.

\clearpage

\begin{deluxetable}{cccr@{.}lr@{.}lcc}
\tablewidth{0pt}
\tablecaption{Dust Region Parameters from the Calibration Fit}
\tablehead{
\colhead{Region}       & 
\colhead{$T$}      & \colhead{$\sigma_T$} &
\multicolumn{2}{c}{$\tau(\nu_0)$} & \multicolumn{2}{c}{$\sigma_\tau$}  & 
\colhead{$\alpha$}     & \colhead{$\sigma_\alpha$} \\
& \colhead{(K)} & \colhead{(K)} 
& \multicolumn{2}{c}{(\ee{-5})} & \multicolumn{2}{c}{(\ee{-5})} & & }
\startdata
\label{tab:DustParams}
Dust 1 &       19.4 & 1.0 & &83  & 0&14 & 1.54 & 0.12 \\
Dust 2 &       15.0 & 0.6 & 1&0  & 0&19 & 2.02 & 0.13 \\
Dust 3 &       14.5 & 0.7 &  11& &   2& & 1.98 & 0.16 \\
Dust 4 &       14.8 & 0.6 & 4&4  & 0&8  & 2.06 & 0.15 \\
Dust 5 &       14.7 & 0.6 & 8&0  & 1&5  & 2.00 & 0.15 \\
\enddata
\end{deluxetable}

\clearpage

\Figure
{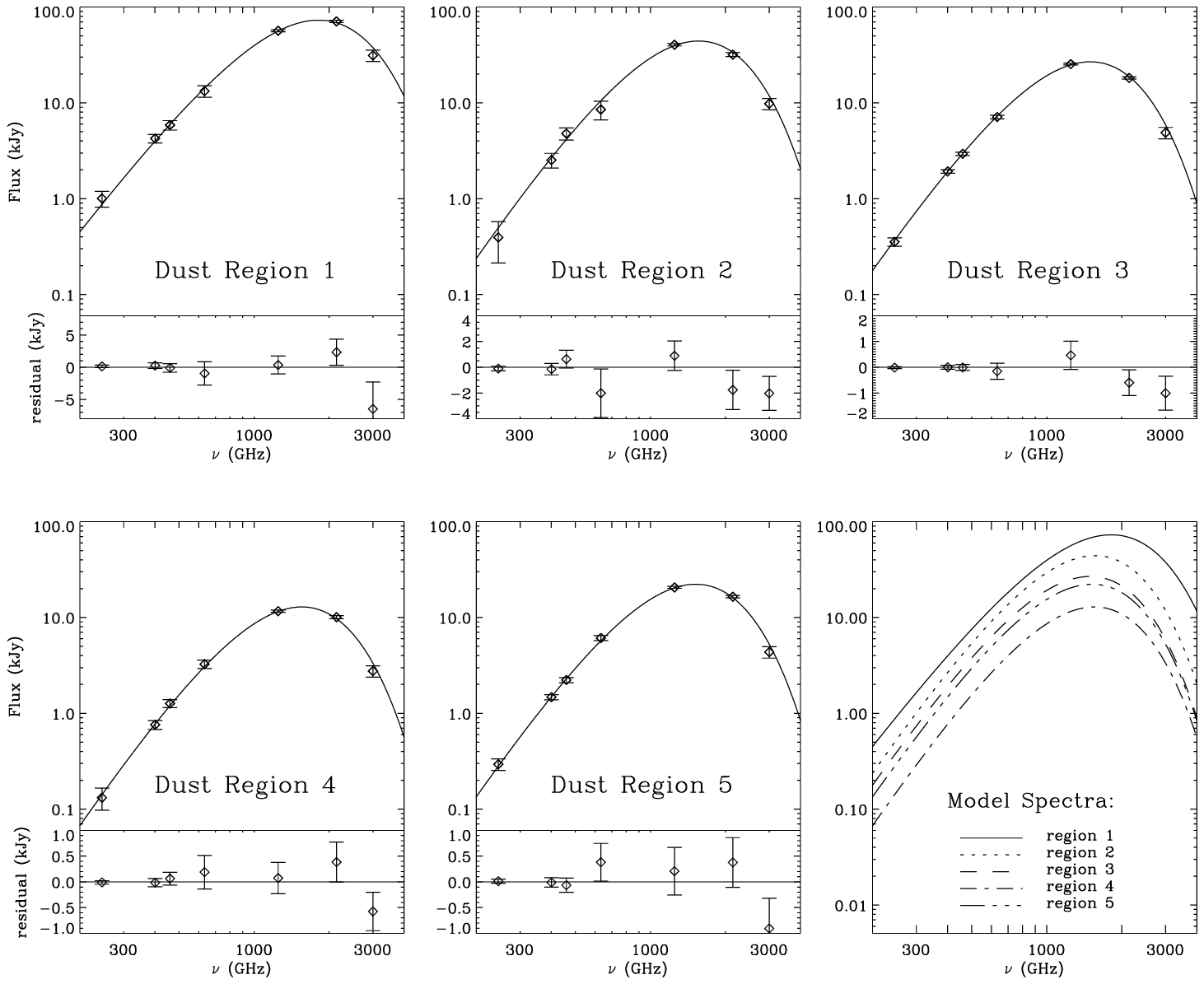}
{Fits to determine the calibration of the TopHat channels at
\chthreebc, \chfourbc, and \chfivebc ~GHz.  The best fit model for
each of dust regions 1 - 5 is shown, together with the residuals from
the fit.  All model spectra are plotted together in the lower right
panel.} {fig:ddcal}

\clearpage

\section{Calibrated Fluxes}
\label{sec:CalibratedFluxes}

Using the calibrations obtained from the fit in Section 
\ref{sec:Calibration} 
we may then compute the calibrated flux of the LMC excluding
30-Doradus, the SMC, and 30-Doradus alone.  The errors in the
calibration are uncorrelated with the flux errors for these regions;
therefore, the covariance of the calibrated flux is the sum of the
covariance of the uncalibrated flux variances and the covariance of
the calibration.  The calibrated differential fluxes and their errors
for these source regions are given in Table
\ref{tab:CalibratedFluxes}.
The errors quoted are the square root of the diagonal elements of the
combined covariance matrix.  We then take these spectra and fit them
to the model of Equation
\ref{eq:DustModel}
for each region separately using the full covariance matrix of their
errors.  The best fit parameters, with errors given by the square root
of the diagonal of the covariance matrix, and the \redchisq~for each
of the fits are given in Table
\ref{tab:FitResults}.
The resulting fitted parameters are highly correlated, particularly
the emissivity and temperature; the correlation matrices for the fits
are given in Table
\ref{tab:CorrelationMatrices}.
The data, best fit spectra, and fit residuals are shown in Figure
\ref{fig:source_spectra}.
We point out that the errors on the parameters as given in Table
\ref{tab:FitResults}
are appropriate $1 \sigma$ errors under the assumption of Gaussian
random errors in the underlying fluxes, and the error for any
parameter individually assumes the other parameters are unconstrained.
When using all parameters together to describe the flux, it is
necessary to consider the correlation as given in Table
\ref{tab:CorrelationMatrices}
in propagating the error; doing so is the only way to capture the full
constraint placed on the spectrum by this measurement.

\clearpage

\begin{deluxetable}{ccrrrrrr}
\tablewidth{0pt}
\tablecaption{
Calibrated fluxes and errors for the LMC, SMC, and 30-Doradus.}
\tablehead{
\colhead{Instrument} & \colhead{Frequency} & \multicolumn{2}{c}{LMC} & 
\multicolumn{2}{c}{30-Dor} & \multicolumn{2}{c}{SMC} \\
& \colhead{(GHz)} & \colhead{$F$ (kJy)} & 
    \colhead{$\sigma_F$ (kJy)} &
    \colhead{$F$ (kJy)} &
    \colhead{$\sigma_F$ (kJy)} &
    \colhead{$F$ (kJy)} &
    \colhead{$\sigma_F$ (kJy)}
}
\startdata
\label{tab:CalibratedFluxes}
TopHat &  \chtwobc & 1.63  & 0.17 & 0.27  & 0.03 & 0.32  & 0.08 \\ 
       &  \chthreebc & 7.93  & 0.59 & 1.26  & 0.10 & 0.95  & 0.19 \\
       &  \chfourbc & 10.57  & 0.89 & 1.85  & 0.16 & 1.62  & 0.29 \\
       &  \chfivebc & 29.66 & 2.98 & 4.54  & 0.47 & 3.20  & 0.81 \\
\tableline
DIRBE  & \dirbetenbc   & 112.3 & 2.4  & 24.99 & 0.57 & 12.07 & 0.54 \\
       & \dirbeninebc  & 177.7 & 3.8  & 46.29 & 1.04 & 20.03 & 0.92 \\
       & \dirbeeightbc & 130.1 & 17.6 & 39.56 & 5.35 & 16.48 & 2.22 \\
\enddata
\end{deluxetable}

\begin{deluxetable}{lrrrrrrrr}
\tablewidth{0pt}
\tablecaption{
Results of model fit to the LMC, SMC, and 30-Doradus.}
\tablehead
{
\colhead{Region} & \colhead{$T$} & \colhead{$\sigma_{T}$} &
          \colhead{$\tau(\nu_0)$} & \colhead{$\sigma_\tau$} &
          \colhead{$\alpha$} & \colhead{$\sigma_\alpha$} &
          \colhead{\redchisq} & \colhead{PTE} \\
\colhead{} & \colhead{(K)} & \colhead{(K)} &
          \colhead{\ee{-5}} & \colhead{\ee{-5}} &
          \colhead{} & \colhead{} &
          \colhead{} & \colhead{}
}
\startdata
\label{tab:FitResults}
LMC  &  25.0 & 1.8 & 0.99 & 0.01 & 1.33 & 0.07 & 5.25/4 & 0.26 \\
SMC  &  29.5 & 2.7 & 0.26 & 0.04 & 0.91 & 0.15 & 1.71/4 & 0.79 \\
30-Dor & 26.2 & 2.3 & 2.1 & 0.4 & 1.50 & 0.08 & 1.65/4 & 0.80 \\
\enddata
\end{deluxetable}

\begin{deluxetable}{llccc}
\tablewidth{0pt}
\tablecaption
{Correlation matrices for the model fits to the LMC, SMC, and 30-Doradus.}
\tablehead
{ \colhead{Region} & \colhead{Parameter} & $T$ & $\tau(\nu_0)$ & $\alpha$ }
\startdata
\label{tab:CorrelationMatrices}
       & $T$           &    1.0000  &   -0.9626  &   -0.6684 \\
LMC    & $\tau(\nu_0)$ &            &    1.0000  &    0.4561 \\
       & $\alpha$      &            &            &    1.0000 \\
\tableline
 &  &                 $T$        & $\tau(\nu_0)$      &    $\alpha$   \\
\tableline
       & $T$           &    1.0000   &   -0.9633  &   -0.6584 \\
30-Dor & $\tau(\nu_0)$ &             &    1.0000  &    0.4430 \\
       & $\alpha$      &             &            &    1.0000 \\
\tableline
& &                  $T$         &    $\tau(\nu_0)$  &    $\alpha$  \\
\tableline
       & $T$           &    1.0000   &  -0.8531   &   -0.6992 \\
SMC    & $\tau(\nu_0)$ &             &   1.0000   &    0.3046 \\
       & $\alpha$      &             &            &    1.0000 \\
\enddata
\end{deluxetable}

\clearpage

\Figure
{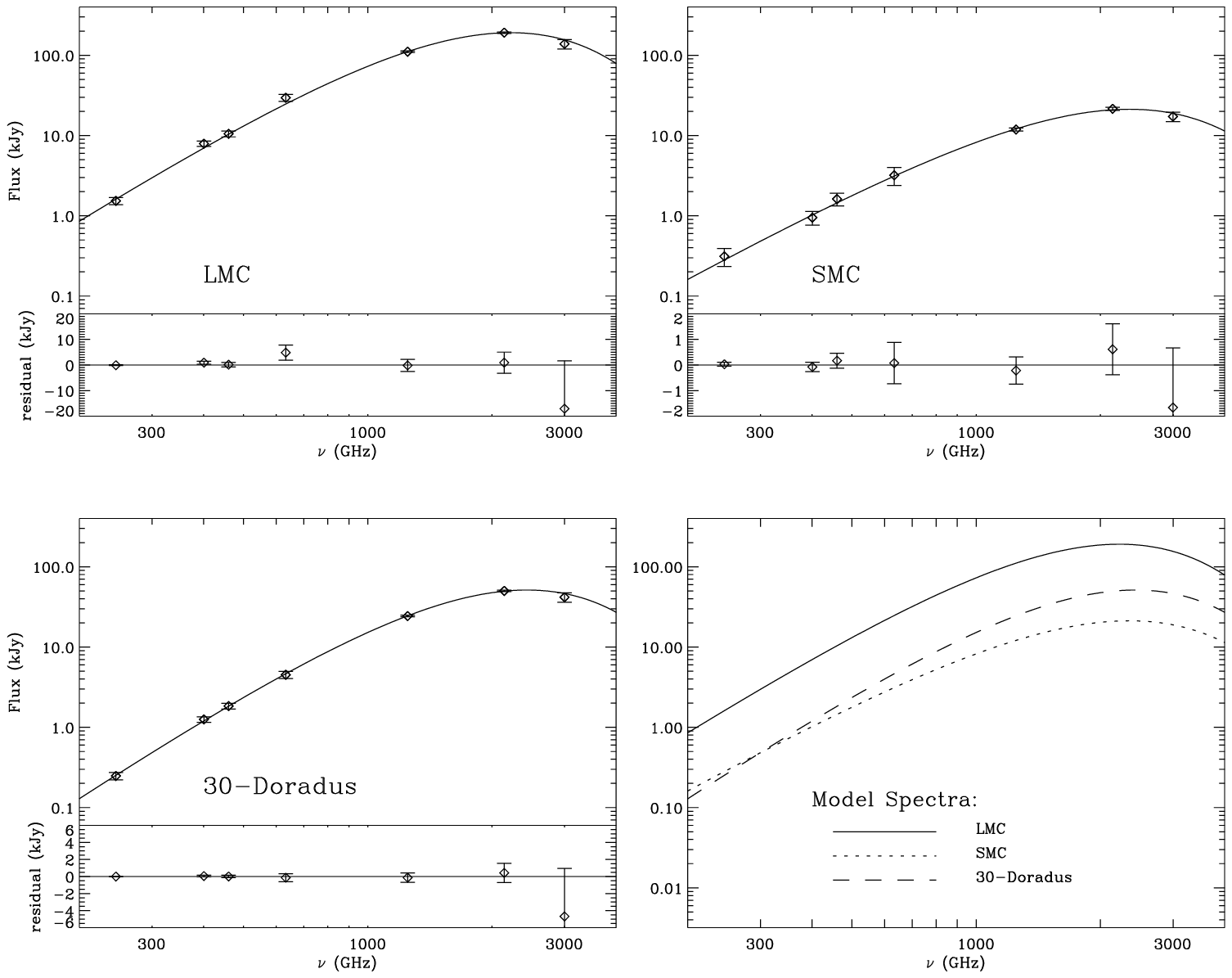}
{Spectrum of calibrated LMC, SMC, and 30-Doradus flux from TopHat and
DIRBE.  All model spectra are plotted together in the lower right
panel.}  {fig:source_spectra}

\clearpage

\section{Discussion}
\label{sec:Discussion}

We have measured the integrated flux relative to the background of the
LMC (minus 30-Doradus), 30-Doradus alone, and the SMC over the
frequency range of \chtwobc~- \dirbeeightbc ~GHz.  The \chtwobc~- 
\chfivebc ~GHz measurements (four bands) are new results derived from
maps of the Southern Polar Cap made by the TopHat telescope, the first
instrument to map this region in these frequency bands with
degree-scale angular resolution.

We have applied calibrations to the TopHat \chtwobc ~GHz data (using
the CMB dipole) and the higher-frequency TopHat channels (using an
interpolation between the TopHat \chtwobc ~GHz and DIRBE measurements
of five galactic dust regions) and reported integrated fluxes for the
LMC, SMC, and 30-Doradus.  The only published measurements of the
integrated flux of these regions in the millimeter/sub-mm continuum
are from
\citet{andreani90}, 
using timestream data from single scans of a $\sim 1\arcdeg$ beam
across the LMC and SMC in two very wide bands ($\Delta \nu_1 \sim 35$
GHz, $\Delta \nu_2 \sim 240$ GHz) with effective band centers of
$\nu_1 = 145$ GHz and $\nu_2 = 260$ GHz.  Their reported surface
brightnesses are compared to those of this work in Table
\ref{tab:SurfaceBrightnessComparison}.
We are unable to account for the orders-of-magnitude discrepancy in
values.  We also show the predictions of the FDS99 model in Table
\ref{tab:SurfaceBrightnessComparison},
and note that the FDS99 predictions are in much closer agreement to
our own.  As mentioned in Section
\ref{sec:Introduction}, 
FDS99 explicitly masks off the Magellanic Clouds in the fit to FIRAS
data from which they derive the global parameters of their model.
Nevertheless, their prediction for the flux in these regions is useful
as an order-of-magnitude guess and is less reliable than their
predictions of galactic dust emission only insofar as the optical
properties of the dust in the Magellanic Clouds differ from those of
the mean high-latitude dust in our galaxy.

\clearpage

\begin{deluxetable}{lccc}
\tablewidth{0pt}
\tablecaption{Comparison of Previous Surface Brightness Values for the 
Magellanic Clouds}
\tablehead{
\colhead{Reference} & \colhead{Center Frequency} & 
\multicolumn{2}{c}{Surface Brightness} \\
 & \colhead{(GHz)} & \colhead{LMC + 30 Dor} & \colhead{SMC}
}
\tablecomments{All surface brightnesses given in units of
10$^{-18}$ W cm$^{-2}$ sr$^{-1}$\mum$^{-1}$}
\startdata
\label{tab:SurfaceBrightnessComparison}
Andreani et al. 1990 & 145      & $198 \pm 59$   & $174 \pm 51$  \\
Andreani et al. 1990 & 260      & $1220 \pm 530$ & $905 \pm 440$ \\
FDS99                & 260      & 75             & 25            \\
This work            & \chtwobc & $24 \pm 2$     & $12 \pm 3$    \\
\enddata
\end{deluxetable}

\clearpage

We have fit the calibrated TopHat and DIRBE measurements of the LMC
(minus 30-Doradus), 30-Doradus alone, and the SMC to an emission model
which is a blackbody times a power-law emissivity and found that a
single temperature and power-law emissivity index fits each of these
regions adequately.  The regions were fit to a two-component model
(with independent temperatures, emissivity power-law indices, and
abundances) as well, and the goodness-of-fit did not improve.  
\citet{stanimirovic00} analyzed DIRBE data using their own foreground 
subtraction process to produce integrated flux from the SMC and found 
that in a ``chi-by-eye'' sense (without a formal fit) their 
fluxes determined from DIRBE band 7, 8, 9, and 10 data (\dirbetenbc 
~GHz ($240 \mum$) - \dirbesevenbc ~GHz ($60 \mum$)) were not 
well fit by a single dust component.  Similarly, \citet{dunne01} found 
that a sample of 32 nearby galaxies observed between 350 GHz and 
\dirbesevenbc ~GHz have spectra more well-described by a two-component 
model than one with a single component.  We also find that if we
extend our frequency coverage by adding DIRBE \dirbesevenbc ~GHz data
(analyzed using the method described in section
\ref{sec:FluxAnalysis}), 
our spectra are no longer fit well by a single component.  But in the
frequency range \chtwobc ~GHz to \dirbeeightbc ~GHz, we find a
single-component fits the emission spectra of the LMC, 30-Doradus, and
the SMC just as well as a two-component model.

If we interpret the results of this fit as a physical description of a
single dust population in the Magellanic Clouds --- rather than a
convenient parameterization --- we can draw a number of conclusions
about the global properties of the dust and the IRF in these regions.
Before doing so, however, we note that other physically plausible
models could produce the observed spectrum.  In fitting FIRAS data to
a greybody emission model,
\citet{reach95} 
point out that the integrated emission from dust with a broad
temperature distribution but a single emissivity power-law index is
difficult to distinguish from greybody emission at a single
temperature and a shallower power-law emissivity.  For example, the
spectra from our three regions can be adequately fit by a distribution
of emitters with $\alpha = 2.0$ (as predicted by the simplest dust
models) and a continuous temperature distribution from
$T_{dust}=T_{cmb}$ up to some maximum temperature $T_{dust}=T_{max}$.
For convenience we assume a power-law dust temperature distribution
$dN/dT \propto T^{-\beta}$.  This gives a model for the flux as
\begin{equation}
\label{eq:TempDistModel}
F_\nu \propto
\int_{T_{cmb}}^{T_{max}} dT \; T^{-\beta} \; B_{\nu}(T) \; (\nu/\nu_0)^{2.0}.
\end{equation}
The \redchisq~ and best-fit $T_{max}$ and $\beta$ from fitting this
model to the flux of the LMC, SMC, and 30-Doradus are shown in Table
\ref{tab:FixsenModel}.

\clearpage

\begin{deluxetable}{lccrr}
\tablewidth{0pt}
\tablecaption{
Results of fit using alternate model of Equation \ref{eq:TempDistModel}.
}
\tablehead
{
\colhead{Region} & \colhead{$T_{max}$} & \colhead{$\beta$} & 
\colhead{\redchisq} & \colhead{PTE}
}
\startdata
\label{tab:FixsenModel}
LMC  &  26.0 & 2.0 & 3.8/4 & .43 \\
SMC  &  29.0 & 3.0 & 1.09/4 & .90 \\
30-Dor & 27.5 & 1.4 & 1.92/4 & .75 \\
\enddata
\end{deluxetable}

\clearpage

This alternate interpretation of our results is bolstered somewhat by
evidence in previous measurements for significant dust temperature
variation within the SMC.
\citet{stanimirovic00} 
report SMC dust temperatures of $23 \kel < T_d < 45 \kel$ based on
{\sl IRAS} $5000 / 3000$ GHz flux ratios.  These values are subject to
some scrutiny as absolute temperatures, due both to the offset and
calibration issues in the {\sl IRAS} data and to the fact that
temperatures derived from $5000$ GHz -- to -- $3000$ GHz flux ratios
are probably most sensitive to the non-equilibrium heating and cooling
of very small grains.  But the observation of variations in this
quantity imply a non-uniform IRF in the SMC, which should result in a
distribution of temperatures in the larger grains as well.

From our single-component model we derive dust temperatures of $(25.0
\pm 1.8)$ K for the LMC (minus 30-Doradus), $(26.2 \pm 2.3)$ K for
30-Doradus alone, and $(29.5 \pm 2.7)$ K for the SMC.  If we instead
assume a broad dust temperature distribution, similar values are
obtained for the hottest dust in each region (Table
\ref{tab:FixsenModel}).  
The general result that the SMC is hotter than the LMC was seen by
\citet{sauvage90}, 
who account for this by the lower observed dust-to-gas ratio in the
SMC, which would imply more UV photons per dust grain.  However, these
same authors find that the ratio of $5000$ GHz to $3000$ GHz emission
is higher in 30-Doradus than in the rest of the LMC, while we find the
temperatures of these two regions to be within $1 \sigma$ of each
other.  One would naively expect 30-Doradus to be hotter than the rest
of the galaxy because it is an active star forming region with plenty
of UV-emitting early-type stars.  As in the SMC, the $5000$ GHz ~/ $3000$ 
GHz \  -- derived relative temperature is sampling the transient
heating of very small grains, but it implies a more intense IRF in
30-Doradus than in the rest of the LMC, which should be detectable in
emission from the larger grains as well.  This situation may be better
explained by the model of a single emissivity power law and a
distribution of grain temperatures.  In this model, the maximum dust
temperature is similar in 30-Doradus and the rest of the LMC, but the
hotter dust makes up a much larger proportion of the total dust in
30-Doradus compared to the rest of the LMC (as evidenced by the
best-fit dust temperature power-law indices in the two regions).  This
makes sense in the context of a simple physical picture in which the
hotter dust component is found in the proximity of hot, early-type
stars.

Using either of the interpretations of our fit results, we can ask the
question ``How hot is the dust in the Magellanic Clouds compared to
the dust in our galaxy?'' FIRAS measured the galactic emission
spectrum in all directions between $30$ GHz and $3000$ GHz, and
several groups have attempted to derive galactic dust parameters using
all or part of the FIRAS coverage.  \citet{reach95} split the FIRAS
coverage into 23 regions in the galactic plane and seven regions above
$|b|=10\arcdeg$ and fit these regions separately to a number of different
models.  They found that in a two-component model with $\alpha = 2.0$
for both populations, the hotter component ranged from $18.6 \kel \le
T \le 24.7 \kel$ in the plane and $16.8 \kel \le T \le 18.3 \kel$ for
the high-latitude regions.  FDS99 fit smoothed {\sl IRAS} and DIRBE
data to FIRAS in regions above $|b|=7\arcdeg$, excluding the Magellanic
Clouds and HII regions in Orion and Ophiuchus.  Their best-fit model
was a two-component greybody fit with floating emissivity power-law
indices for each component, and the mean temperatures of the two
components were $\langle T_1 \rangle = 9.4 \kel$ and $\langle T_1
\rangle = 16.2 \kel$.  We find that the temperature of the dust in the
Magellanic Clouds --- or the temperature of the hottest dust component
in each of the three observed regions --- is on the high side of all
of these galactic measurements.  The temperature we derive for the SMC
is significantly higher than any component from the galactic
measurements, while the temperatures we derive for the LMC and
30-Doradus are comparable to the hottest regions seen by
\citet{reach95} in the plane of our galaxy.

We derive effective emissivity power-law indices of $1.33 \pm 0.07$
for the LMC (minus 30-Doradus), $1.50 \pm 0.08$ for 30-Doradus alone,
and $0.91 \pm 0.15$ for the SMC.  In fitting FIRAS data at high
galactic latitudes,
\citet{reach95} 
found a similar range of effective power-law indices ($0.92 < \alpha <
1.60$) for a one-component fit with $\alpha$ as a free parameter.
They find slightly better fits with a two-component model and fixed
$\alpha = 2.0$ but no improvement when they add a slightly broadened
uniform temperature distribution.
\citet{pollack94}, 
in modeling IR dust emission from circumstellar accretion disks at
$\sim 100 \kel$ predict an index of $\sim 1.5$ below $500$ GHz, going
over to an index of $\sim 2.6$ at higher frequencies.  According to
the authors, this is due to the changes in the relative contribution
of so-called astronomical silicates --- which the authors predict
should have an index of $1$ --- and organic species.  In fitting
extrapolated and smoothed {\sl IRAS} and DIRBE data to FIRAS
measurements, FDS99 found that a two-component model close to this
prediction ($\alpha_1 = 1.67$, $\alpha_2 = 2.7$, with equal power
radiated by the two components in the area of $500$ GHz) was the best
fit to the high-latitude emission in our galaxy.  However, in the
three regions we observed we found no evidence for a change in the
emissivity power-law index over the observed frequencies (\chtwobc~-
\dirbeeightbc ~GHz).

Finally, we note that if we assume we have sampled a single component
of isothermal dust in each region, we can calculate a total mass for
that component of dust using the observed optical depth at a
particular frequency and a value for the opacity of the observed dust
component at that frequency.  Of course, published values for
low-frequency opacities of likely candidates for astrophysical dust
vary widely, and this contribution to the uncertainty in the inferred
dust mass dominates the formal uncertainty on the measured optical
depth.  For example, we can assign an absolute dust opacity to our
observed regions by comparing the observed frequency dependence of
dust opacity in each region we observed with the measured frequency
dependence of various grains in
\citet{agladze96} 
and attempting to find a grain species for which their laboratory
measurements match our observations.  In doing so, we find that at the
temperatures we infer for the observed regions, the best matches were
amorphous $\mathrm{MgO} \cdot 2\mathrm{SiO}_2$ with measured $\alpha
\sim 1.1$ at 23 K for the SMC, and amorphous $2\mathrm{MgO} \cdot 
\mathrm{SiO}_2$ with measured $\alpha \sim 1.7$ at 25 K for the LMC 
and 30-Doradus.  The measured absolute opacities of these species at
300 GHz were 3.29 cm$^2$ g$^{-1}$ and 1.04 cm$^2$ g$^{-1}$.  Using
these values and the $300$ GHz optical depth from our best-fit
greybody models, we derive integrated dust masses of $(6.2 \pm 1.1)
\times 10^5 M_{\odot}$ for the LMC minus 30-Doradus, $(9.8 \pm 2.1)
\times 10^4 M_{\odot}$ for 30-Doradus alone, and $(4.0 \pm 0.9) \times
10^4 M_{\odot}$ for the SMC.  (These values assume distances of $(49
\pm 2) ~\mathrm{kpc}$ to the LMC and $(60 \pm 3) ~\mathrm{kpc}$ to the
SMC
\citep{westerlund97}.)
The value thus obtained for the SMC is inconsistent with the value of
$(1.8 \pm 0.2) \times 10^4 M_{\odot}$ obtained in
\citet{stanimirovic00}.  
However, these authors assumed an absolute dust opacity at $\nu=3000$
GHz of $41$ cm$^2$ g$^{-1}$ (which is close to the DL84 prediction for
both graphite and silicate species).  And if we apply this value to
the $3000$ GHz optical depth from our best-fit SMC model, we obtain a
value of $(2.6 \pm 0.8) \times 10^4 M_{\odot}$, which is consistent
with the
\citet{stanimirovic00} 
result.

However, the greatest uncertainty on this method of calculating dust
masses stems from the possibility that the strict physical
interpretation of our results is not the correct one.  For example, if
in fact there is a distribution of temperatures in the SMC with
$T_{max} = 30 \kel$ and $dN/dT \propto T^{-3}$, the inferred dust mass
will be an order of magnitude higher due to the ``hidden'' cold dust
component.  The possibility of hidden cold dust was also recognized in
\citet{stanimirovic00}, whose dust mass calculation was based on 
temperatures derived from $\nu \ge 3000$ GHz data only and was therefore 
insensitive to dust below $\sim 20 \kel$.

\section{Conclusions}
\label{sec:Conclusions}

We have demonstrated that for each of three extragalactic regions (the
Large Magellanic Cloud excluding 30-Doradus, 30-Doradus alone, and the
Small Magellanic Cloud), the integrated flux in seven frequency bands
between \chtwobc ~GHz and \dirbeeightbc ~GHz is well described by a
simple, one-component greybody emission model with power-law
emissivity.  Though it is difficult to obtain robust information about
global properties of these regions (such as total dust mass) from
these results, the results are intriguing as a potential road map for
characterizing the emission in these frequencies across many types of
galaxies.  If the statistics of emission from extragalactic
environments in these frequencies could be accurately described by
distributions in a few simple parameters, it would be a boon to groups
seeking to probe structure formation by associating the correlation
properties of the Cosmic Infrared Background with the distribution of
dusty protogalaxies at various redshifts (c.f.
\citet{kceh01}).  
And knowledge of the ``typical'' emission in these frequencies from
extragalactic environments --- and the variation from galaxy to galaxy
--- is critical for efforts to probe the earliest collapsed structures
through source counting (c.f. 
\citet{blain99a}), 
because only photometric redshifts will be available for most of the
sources.

We are able to put useful constraints on this small set of model
parameters for the regions we observe because we have a relative
wealth of frequency coverage: seven bands for a three-parameter model.
The instruments that give this particular combination of bands (TopHat
and \COBE/DIRBE) have angular resolution that is appropriate for
obtaining integrated fluxes from the Magellanic Clouds, but not for
extending this technique to a wider sample of galaxies.  The {\sl
IRAS} Bright Galaxy Sample is a useful source of information at $\nu
\ge 3000$ GHz, and data from the SCUBA instrument has already added
points at $350$ GHz and $670$ GHz for many of these galaxies
\citep{dunne01}.  
To obtain constraints of the nature of those we have placed on the
Magellanic Cloud flux, however, it will be critical both to augment
the sub-mm coverage (which can be achieved with several existing and
planned ground-based instruments) and to add points in frequency bands
analogous to the DIRBE \dirbeninebc ~GHz and \dirbetenbc ~GHz bands.
This area of the spectrum is inaccessible from the ground, so
measurements there will have to come from balloon, high-altitude
aircraft, or satellite missions.  The existing Infrared Space
Observatory (ISO)\footnote{http://www.iso.vilspa.esa.es/} dataset and
anticipated results from the Space Infrared Telescope Facility
(SIRTF)\footnote{http://sirtf.caltech.edu/} and the Stratospheric
Observatory For Infrared Astronomy
(SOFIA)\footnote{http://sofia.arc.nasa.gov/} are candidates for
filling in this gap.  In addition, at lower frequencies, the Wilkinson
Microwave Anisotropy Probe (WMAP)\footnote{
http://lambda.gsfc.nasa.gov/product/map/m\_products.html } data could
be combined with the TopHat and DIRBE data to determine at what
frequency other sources of diffuse emission begin to dominate over
thermal dust.

\acknowledgements

We would like to thank the National Scientific Balloon Facility for
balloon launch services.  We would also like to thank Julian Borrill
and acknowledge the use of resources of the National Energy Research
Scientific Computing Center.  We are also grateful to colleagues at
the Danish Space Research Institute and Niels Bohr Institute for their
contributions in subsystem development, testing and fabrication and
flight support, especially Rene Kristensen and Per Rex Christensen.
This research was supported by the NASA Office of Space Sciences and
by NSF grant OPP-9619374 and NASA grants NAG5-11443 and NGT5-86/SM.

\appendix
\section{Calculation of Color Corrections}
\label{app:ColorCorrections}

If an experiment has finite bandwidth ($t(\nu) \ne \delta(\nu -
\nu_c)$), to report a source surface brightness at a single
frequency, one must assume a source spectrum.  The power detected from
that source is assumed to be
\begin{equation}
P_{in} = \eta \; A \Omega \; \int I_0(\nu) t(\nu) d \nu
\end{equation}
(Here $\eta$ and $A \Omega$ are the optical efficiency and throughput
of the instrument, $I_0(\nu)$ is the nominal (assumed) surface
brightness of the source, and $t(\nu)$ is the bandpass transmission
normalized to $1.0$ at its peak.)  The effective band center $\nu_c$
is usually chosen such that
\begin{equation}
I_0(\nu_c) \simeq \frac{\int I_0(\nu) t(\nu) d \nu}
{\int t(\nu) d \nu} 
\end{equation}
The band centers for TopHat are calculated assuming a Rayleigh-Jeans
(RJ) source spectrum
\begin{eqnarray}
I_0(\nu_c) & = & I_{RJ}(\nu_c) \\
\nonumber & = & \tau \; 2 k T \; \frac{\nu_c^2}{c^2},
\end{eqnarray}
where $\tau$ is the optical depth of the source and $k$ is Boltzmann's
constant.  Since the detected power is assumed to be
\begin{equation}
P_{in} = \eta \; A \Omega \; \int \tau \; 2 k T \; \frac{\nu^2}{c^2}
\; t(\nu) d \nu,
\end{equation}
we can write
\begin{equation}
I_0(\nu_c) = \frac{P_{in}}{\eta \; A \Omega \int \nu^2 \; t(\nu) d \nu} \; \nu_c^2
\end{equation}
Now if we assume a different source spectrum, for example a greybody
with power-law emissivity, the assumed input power is
\begin{eqnarray}
P_{in} & = & \eta \; A \Omega \; \int I_{GB}(\nu) t(\nu) d \nu \\
\nonumber & = & \eta \; A \Omega \; \int \tau(\nu_0) (\nu / \nu_0)^\alpha B_\nu(T) t(\nu) d \nu
\end{eqnarray}
and the source spectrum inferred from the detected power is
\begin{eqnarray}
I_{GB}(\nu_c) & = & \tau(\nu_0) (\nu_c / \nu_0)^\alpha B_{\nu_c}(T) \\
\nonumber & = & \frac{P_{in}}{\eta \; A \Omega \int \nu^\alpha B_\nu(T) t(\nu) d \nu} \; \nu_c^\alpha B_{\nu_c}(T) \\
\nonumber & = & \frac{\int \nu^2 t(\nu) d \nu}{\int \nu^\alpha B_\nu(T) t(\nu) d \nu} \; \nu_c^{\alpha-2} B_{\nu_c}(T) \; I_{RJ}(\nu_c) \\
\nonumber & \equiv & \frac{I_{RJ}(\nu_c)}{K}.
\end{eqnarray}
This defines the color correction $K$ to apply to the reported TopHat
flux from a source if the source is assumed to have a greybody
spectrum with power-law emissivity.

We make a similar calculation for DIRBE, for which the band centers are 
computed assuming a spectrum with $\nu I(\nu)$ constant.  In this case, 
the correction is given by 
\begin{eqnarray}
K &=& \frac{\nu_c^{-1}}{\int \nu^{-1} t(\nu) d \nu} \left[\frac{\tau(\nu_0) (\nu_c / \nu_0)^\alpha B_{\nu_c}(T)}{\int \tau(\nu_0) (\nu / \nu_0)^\alpha B_{\nu}(T) t(\nu) d \nu} \right]^{-1} \\
\nonumber &=& \frac{\int \nu^\alpha B_{\nu}(T) t(\nu) d \nu}{\int \nu^{-1} t(\nu) d \nu} \; \nu_c^{-(\alpha+1)} B_{\nu_c}(T)
\end{eqnarray}
For an arbitrary experiment with bandpass $t(\nu)$ that reports its
surface brightness measurements assuming a spectrum $I_0(\nu)$, the
surface brightness assuming a different source spectrum $I_1(\nu)$ is
given by
\begin{eqnarray}
I_1(\nu_c) &=& I_0(\nu_c) / K \\
\nonumber &=& I_0(\nu_c) \frac{I_1(\nu_c)}{\int I_1(\nu) t(\nu) d \nu} \left[\frac{I_0(\nu_c)}{\int I_0(\nu) t(\nu) d \nu} \right]^{-1}
\end{eqnarray}

\end{document}